\documentclass{an}
\usepackage{graphicx}
\usepackage{times}
\begin{document}

\Pagespan{1}{}
\Yearpublication{2010}%
\Yearsubmission{2010}%
\Month{}%
\Volume{}%
\Issue{}%
\DOI{DOI}%

\def\ds{$\delta$\,Sct}
\def\gdor{$\gamma$\,Dor}
\def\Tf{T$_{\mathrm{eff}}$}
\def\lg{log\,$g$}
\def\vsini{$v\sin i$}
\def\kms{$\rm{kms^{-1}}$}
\def\cd{\,$\rm{d}^{-1}$}
\def\corot{CoRoT}
\def\lcs{light curves}

\title{\gdor\ and \gdor\ - \ds\ Hybrid Stars In The CoRoT\thanks{The CoRoT space mission was developed and is operated by the French space agency CNES, with participation of ESA's RSSD and Science Programmes, Austria, Belgium, Brazil, Germany, and Spain. } LRa01}

\author{M. Hareter\inst{1}\fnmsep\thanks{Corresponding author:
  \email{markus.hareter@univie.ac.at}\newline}
\and P. Reegen\inst{1}
\and A. Miglio\inst{2}
\and J. Montalb\'an\inst{2}
\and A. Kaiser\inst{1}
\and I. Dek\'any\inst{3}
\and E. Guenther\inst{4}
\and E. Poretti\inst{5}
\and P. Mathias\inst{6}
\and W. Weiss\inst{1}
}
\institute{
Institut f\"ur Astronomie, Universit\"at Wien, T\"urkenschanzstrasse 17, A-1180 Vienna, Austria
\and Institut d'Astrophysique et de Geophysique Universit\'e de Li\`ege, All\'ee du 6 Ao\^ut 17, B 4000 Li\`ege, Belgium
\and Konkoly Observatory, PO Box 67, H-1525 Budapest, Hungary
\and Instituto de Astrofísica de Canarias C/ V\'ia L\'actea, s/n E38205 - La Laguna (Tenerife), Spain
\and INAF -- Osservatorio Astronomico di Brera, via E. Bianchi 46, 23807 Merate, Italy
\and Observatoire Midi Pyrenees, Laboratoire d'Astrophysique, 57 Avenue d'Azereix,F-65000 Tarbes, France
}
\received{}
\accepted{}
\publonline{}


\abstract{ A systematic search for \gdor\ and \gdor\ - \ds\ hybrid pulsators was conducted on the \corot\ LRa01 Exo-archive yielding a total of 418 \gdor\ and 274 hybrid candidates. After an automatic jump correction 194 and 167 respectively, show no more obvious jumps and were investigated in more detail. For about 25\% of these candidates classification spectra from the Anglo-Australian Observatory (AAO) are available. 
The detailed frequency analysis and a check for combination frequencies together with spectroscopic information allowed us to identify I) 34 \gdor\ stars which show very different pulsation spectra where mostly two modes dominate. Furthermore, a search for regularities in their oscillation spectra allowed to derive recurrent period spacings for 5 of these \gdor\ stars.
II) 25 clear hybrid pulsators showing frequencies in the \gdor\ and \ds\ domain and are of A-F spectral type. 
  }
\maketitle

\section {Introduction}
\gdor\ stars are late A- to early F-type stars on the cool border of the \ds\ Instability Strip (IS) in the Hertzsprung-Russell diagram.
\gdor\ stars pulsate in high-order g-modes, with typical periods between 0.3 and 3 days, whose driving mechanism was attributed to the flux blocking mechanism at the bottom of the convective envelope (Guzik et al. 2000).

\ds\ stars pulsate in contrast to \gdor\ stars with low-order p-modes with typical frequencies between 5 and 50\cd\ and excited by the $\kappa$-mechanism in the HeII ionization region. Since the \ds\ stars are studied for decades, the theory is more advanced than for \gdor\ stars. The instability strips of the \gdor\ and \ds\ stars overlap at the cool region of the \ds\ instability strip (e.g. Warner et al. 2003, Dupret et al. 2004). It is natural to expect stars showing both types of pulsation and theoretical computations (Dupret et al. 2004) predict in fact hybrid pulsators showing simultaneously low-order p- and high-order g-modes.

A few discoveries of \gdor/\ds\ hybrids have been reported: HD 8801 (Henry \& Fekel 2005, Handler 2009) and HD 49434 (Uytterhoeven et al. 2008) from ground based observations; HD 114389 and BD+18 4914 (King et al. 2006; Rowe et al. 2006) from space photometry with MOST (Micro-Oscillations of STars, Walker et al. 2003). A first interpretation of KEPLER data (Grigahc\`ene et al. 2010) suggests that there are practically no pure \ds\ or \gdor\ stars.

In this paper we report on the search of \gdor\ and \gdor\ / \ds\ hybrid pulsators in the more than ten thousand light curves obtained in the Exo-channel of the satellite \corot\ (Auvergne et al. 2009). The main advantages of this study with respect to the above mentioned are the high quality of the light curves thanks to the negligible stray light contamination, the long duration of the observations (131 days) with a short cadence of 512 s (at maximum) which allows us to cover frequencies up to 84\cd\ with a precision of 0.008\cd\ for a large sample of stars.


\section {Data Reduction and Frequency analysis}
In spite of the high quality of \corot\ data, minor instrumental effects and the consequences of charged particle hitting the detector lead to trends and jumps in the \lcs\ what makes an additional processing of the \corot\ archive data (N2 data) necessary. 

All N2 light curves were extracted from the archive. In the case of chromatic light curves the fluxes of the individual channels were summed. After the extraction the light curves were automatically detrended and destepped employing a two sample t-test. This test computes the significance of the difference in mean values for two given samples. The two samples contain in our case 50 data points before and the same after a possible step. If the significance exceeds a certain threshold, a jump is assumed and the mean values of both samples are adjusted. (Kaiser et al. in preparation) The rejection policy of the outliers follows three sigma clipping relative to a running average and the corresponding Discrete Fourier Transformation (DFT) was calculated. Each individual light curve and DFT was inspected by eye and those with signal in the \gdor\ and/or \ds\ range were selected. 

A total of 11408 light curves was checked and 692 light curves were selected for further investigation. They split as follows: i) 418 pure \gdor\ / SPB (B-type pulsators with oscillation modes in the same frequency domain) candidates with no obvious signal above 5 c/d, 194 out of these 418 show smooth light curves after our jump-removing pipeline. ii) 274 hybrid candidates with both signal in the low frequency and in the \ds\ regime. 167 out of these 274 hybrid candidates have smooth light curves yielding to a total of 361 (194 \gdor\ / SPB + 167 hybrid candidates) light curves to analyze.
The frequency analyzes were carried out using the software SigSpec (Reegen 2007), where the upper limit of the frequency range for the \gdor\ candidates was set to 5 \cd\ and 84 \cd (Nyquist frequency) for the hybrid candidates.

For 4112 stars in the \corot\ LRa01 classification spectra from AAO are available, which allow to distinguish between \gdor\ and SPB stars. 
For 55 of the 194 \gdor\ /SPB candidates classification spectra are available. An automatic classification based on the line depth ratio of the Ca K and H lines (at 3933 and 3964\AA)
leads to the identification 34 \gdor\ stars and 18 SPB stars, and three A-type stars. 

Out of the 167 hybrid candidates 66 have classification spectra, where 20 out of these spectra indicate early F-type stars, 26 have intermediate strength of the Ca K line, indicating mid to late A-type stars and 20 show weak lines indicating hot stars (B to mid A type). This rough classification was done based on the Ca K line and a more detailed analysis of the spectra is currently being developed. 

\section {\gdor\ and Period Spacings}
\begin{figure}
{\includegraphics[width=80mm,height=60mm]{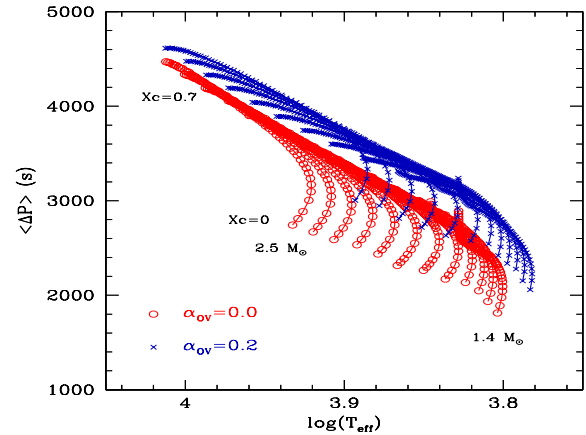}
\caption{Theoretical period spacings of $\ell$ = 1 high-order g-modes in main sequence models as a function of \Tf. Each line connects models of the same mass. } }
\label{fig-theo-spacing}
\end{figure}

The sound detection of a recurrent period spacing would allow detailled inferences on the properties of the near-core region of intermediate-mass stars. (Fig. \ref{fig-theo-spacing} and Miglio et al. 2008). Therefore, a search for regular patterns in the periodograms of the 34 bona fide \gdor\ stars was performed. Three methods were applied to the significance spectrum (Reegen 2007) vs. period: the autocorrelation function, a modified autocorrelation employing products of significances for a test spacing and a similar method using mean significances instead of products to overcome the problem of missing modes.

The autocorrelation method is widely used for this purpose and we do not explain it in more detail here. The modified autocorrelation function is basically comparable to the comb response (Kjeldsen et al. 1995) used for solar like stars, which pulsate in high order p-modes with equidistant frequency patterns. For a given test spacing centered on the highest peak the significances two orders to the lower and two orders to the higher periods are multiplied including the significance of the highest peak (5 multiplicands spaced with the test spacing). This method fails, if a mode is not excited, because one close to zero significance brings the whole product close to zero. The advantage of this method is that the detection of a random spacing of two dominant modes can be nearly excluded.
To overcome the problem of the missing mode, the same method but using averaged significances is used. The advantage of one method is the disadvantage of the other, hence both can be combined to detect spacings even with missing modes.

\begin{figure}
{\includegraphics[width=80mm,height=55mm]{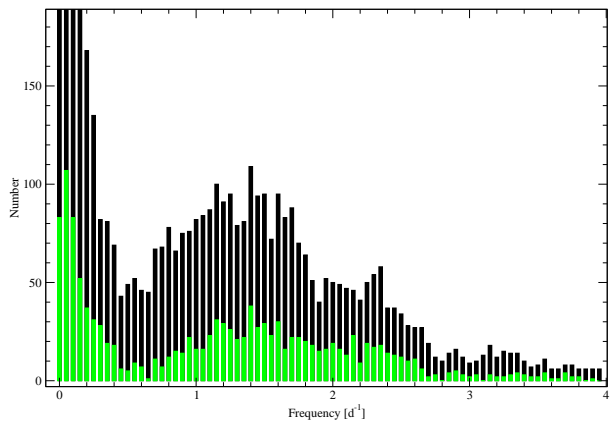}
\caption{Histogram of frequencies detected in all 194 \gdor\ / SPB candidates (black bars) compared to the 34 confident \gdor\ stars (green bars).} }
\label{fig-gdor-histo}
\end{figure}

These methods were applied on the 34 \gdor\ stars and yielded 5 cases were a spacing could be found (Table \ref{tab-spacing}) in the expected range of test spacings (0.001 to 0.05 d). In the other 29 cases no clear spacing was detected. The search for constant spacings is dominated and biased towards the most prominent peaks, whereas peaks at moderate or low significances are underestimated. This disadvantage can be overcome by a more sophisticated search which incorporates a prewhitening of the most dominant peaks. Such a technique is currently being developed, but no results are available at this time.
Fig \ref{fig-gdor-histo} compares the distribution of the \gdor\ candidates (194 stars - likely containing also SPB stars, black bars) to the confirmed cases (34 stars, green bars). No significant deviation is detected. The high number of low frequencies could be due to uncorrected artifacts or long term variations of the stars.

\begin{table}
\caption{Results of the search for regular period spacings.}
\label{tab-spacing}
\begin{tabular}{cccc}\hline
CoRoT ID & Main Period & Spacing & Beating Period\\ 
 & (d) & (s) & (d) \\
\hline
0102603266 & 0.4107 & 648  & 6.570 \\
0102608070 & 0.2908 & 873  & 8.140 \\
0102724193 & 0.4748 & 1391 & 6.169 \\
0102732872 & 0.9178 & 2246 & 11    \\
0102795873 & 0.5648 & 2376 & 4.523 \\
\hline
\end{tabular}
\end{table}

\section {Hybrid Stars}
\begin{figure}
{\includegraphics[width=80mm,height=55mm]{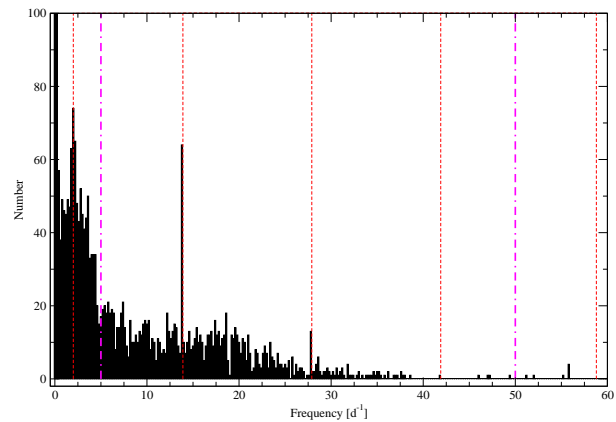}
\caption{Histogram of frequencies detected in all 167 hybrid candidates. The CoRoT orbit frequency and its harmonics and the artifact at 2\cd\ are marked by dashed vertical lines. The dashed-dotted magenta lines indicate the \ds\ frequency domain. } }
\label{fig-hybrids}
\end{figure}

The frequency solutions determined using SigSpec were \\ checked for linear combinations of frequencies consisting of up to 4 terms ($f_c$ = $a f_1  + b f_2 + c f_3 + d f_4$, $a .. d$ being integer numbers, $f_c$ denoting the combination frequency and $f_{1..4}$ the independent frequencies). Out of 167 candidates, 54 stars are identified as hybrid stars, where in 25 cases AAO classification spectra are available and indicate A-F type stars. Among the remaining 113 stars, 50 are still considered candidates because a restrictive combination frequency search removes either the presumed g- or p-modes. The residual sample (63 objects) contain 17 definitely hot stars, outside the \ds\ instability strip according to their classification spectra, 12 probably hot stars according to their pattern of pulsation frequencies, and 34 objects without any indication of hybrid pulsation after checking with combination frequencies. Fig. \ref{fig-hybrids} shows the distribution of frequencies of the 167 hybrid candidates. The \corot\ orbit harmonics are indicated by the vertical dashed lines as well as the artifact at 2\cd\ caused by the South Atlantic Anomaly.

\section {Conclusions}
From the analysis of the LRa01 exo-field light curves we confidently detect a sample of 34 \gdor\ stars and 25 \gdor\ / \ds\ hybrid pulsators. For those stars the analysis of the classification spectra is in progress and we will soon be able to give effective temperatures. With additional spectra or narrow-band photometry it would be possible to confirm many cases of the still remaining 50 hybrid candidates. Among the stars which are not yet analyzed because of their bad light curves we expect to find the same order of \gdor\ and hybrid stars. The number of the known \gdor\ stars will be doubled and the number of the hybrids will be multiplied by a factor of 6.
The histogram shown in Fig. \ref{fig-gdor-histo} indicates that the most frequent frequencies lie between 1 and 2 c/d - except for the very low frequencies, which are most probable not due to g-mode pulsation. The fact, that the search for recurrent period spacings yielded only in 5 cases results confirms that the pulsation behavior of the \gdor\ stars is more complex than predicted by the first order asymptotic theory. First of all, the effect of rotation on these low frequency modes would probably destroy the regular period spacing (Ballot et al. 2009; Bouabid, Dupret \& Grigahc\'ene this proceedings). 
Furthermore, we did not take into account deviations of the recurrent spacings due to the presence of chemical composition gradients that are built in the course of the stellar evolution at the border of the convective core (Miglio et al. 2008).
The distribution of frequencies for the hybrid stars shows that only in a few cases there are p-modes with frequencies higher than 30 c/d. 

\acknowledgements
This research is based on CoRoT data. The CoRoT (Convection Rotation and planetary Transits) space mission, launched on December 27th 2006, has been developed and is operated by CNES, with the contribution of Austria, Belgium, Brazil, ESA, Germany and Spain. We wish to thank the CoRoT team for the acquisition and reduction of the CoRoT data.
JM acknowledges financial support from the Prodex-ESA Contract Prodex 8 COROT (C90199). AM is a postdoctoral researcher of the 'Fonds de la recherche scientifique' FNRS, Belgium.
EP acknowledges the financial support of the Italian ESS project (contract \\ ASI/INAF/I/015/07/0, WP~03170). EG acknowledges the support by the Optical Infrared Coordination network (OPTICON), supported by the Research Infrastructures Programme of the European Commissions in its Sixth Framework Programme for the AAT observations.

\end{document}